\documentclass[twocolumn,superscriptaddress,aps]{revtex4}
\usepackage{amsmath}
\usepackage{amssymb}
\usepackage{amsfonts}
\usepackage{graphicx}
\usepackage{bbold}
\usepackage{bm}
\usepackage{bm}
\usepackage{cancel}
\usepackage{times,float}
\usepackage{graphicx}
\usepackage[dvipsnames,svgnames]{xcolor}
\usepackage{hyperref}
\usepackage{comment}
\hypersetup{colorlinks=true, linkcolor=Blue, citecolor=PineGreen,urlcolor=Blue}
\usepackage{multirow}
\usepackage{ulem}
\usepackage{color,epsfig}
\usepackage{bm}
\usepackage{slashed}
\usepackage{subfigure}
\usepackage{feynmp}
\usepackage{etoolbox}
\usepackage{orcidlink}
\usepackage{braket}
\usepackage{cancel}
\hypersetup{
    colorlinks=true,
    linkcolor=Blue,
    citecolor=Blue,
    urlcolor=Blue
}

\newcommand{\bea}{\begin{eqnarray}}
\newcommand{\eea}{\end{eqnarray}}
\newcommand{\be}{\begin{equation}}
\newcommand{\ee}{\end{equation}}

\renewcommand\vec{\bm}

\begin{document}

\title{Leading order quark--antiquark--$W$-boson vertex in the presence of a magnetic field}
\author{Alejandro Ayala~\orcidlink{0000-0003-3929-9209}}
\email{ayala@nucleares.unam.mx}
\affiliation{Instituto de Ciencias
Nucleares, Universidad Nacional Aut\'onoma de M\'exico, Apartado
Postal 70-543, CdMx 04510,
Mexico}
\author{Alejandro L\'opez~\orcidlink{0009-0005-3672-4312}}
\email
[Corresponding author: ]{aleh.lopezg@gmail.com}
\affiliation{Instituto de Física, Universidade de São Paulo, Rua do Matão, 1371, CEP 05508-090, São Paulo, SP, Brazil}
\author{Ana Julia Mizher
\orcidlink{0000-0001-9502-9815}}
\email{anamizher@gmail.com}
\affiliation{Instituto de Física, Universidade de São Paulo, Rua do Matão, 1371, CEP 05508-090, São Paulo, SP, Brazil}
\author{Manuel Emiliano Monreal Cancino~\orcidlink{0009-0007-4313-2758}}
\email{emilianomonreal@ciencias.unam.mx}
\affiliation{Facultad de Ciencias, Universidad Nacional Aut\'onoma de M\'exico, Apartado Postal 50-542, CdMx 04510, Mexico}
\author{Norberto Scoccola~\orcidlink{0000-0001-9131-4029}}
\email{scoccola@gmail.com}
\affiliation{Departamento de F\'isica Te\'orica,
Comisi\'on Nacional de Energ\'ia At\'omica, Av. Libertador 8250, (1429) Buenos Aires, Argentina}
\affiliation{CONICET, Rivadavia 1917, (1033) Buenos Aires, Argentina}


\begin{abstract}

We compute the leading order modification to the quark-antiquark-$W$-boson vertex in the presence of a constant and uniform magnetic field, specifically for the case where the quark and antiquark are the $u$ and $\bar{d}$, respectively. We find the selection rules for transitions where the $u$, $\bar{d}$, and $W$-boson occupy arbitrary Landau levels. We show that, for the particular case where the initial particles each occupy the lowest Landau level, the $W^+$ is also produced in the lowest Landau level with a single polarization aligned with the magnetic field. We also illustrate the general findings by computing the case of transitions between low lying Landau levels and show the correspondence between the polarization vectors for the $W$-boson and the polarization states for the quark and the antiquark.

\end{abstract}

\maketitle

\section{Introduction}

In recent years, a great deal of attention has been devoted to understanding how magnetic fields modify particle properties. In particular, it has been shown that particle masses and interaction strengths become field dependent, evolving with the magnetic field intensity~\cite{Gusynin:1999pq,Mueller:2014tea,Avancini:2015ady,Avancini:2016fgq,Avancini:2018svs,Ayala:2015qwa,Carlomagno:2022inu,Ayala:2018zat,Avancini:2021pmi,Coppola:2018vkw,Coppola:2023mmq,Li:2016dta,Ayala:2006sv,Ayala:2019akk,Ayala:2021lor,Ayala:2020muk,Ayala:2020dxs,Rojas:2008sg,Castano-Yepes:2022luw,Castano-Yepes:2023brq,Castano-Yepes:2024ltr,Moreira:2022dwo,Castano-Yepes:2024ctr,Wang:2026xsm,Ayala:2025qag,Ayala:2026eja,Adhikari:2024bfa}. The presence of an external magnetic field profoundly modifies the symmetry properties of a relativistic quantum field theory. By selecting a preferred spatial direction, the magnetic field explicitly breaks the isotropy of space and reduces the original Lorentz symmetry to the subgroup that preserves the field orientation. The external magnetic field also explicitly  breaks those discrete symmetries that reverse the direction of the magnetic field, such as charge conjugation and time reversal individually, while preserving their appropriate combinations (e.g. $CP$) and naturally separates the four-dimensional space-time into longitudinal and transverse subspaces with respect to the magnetic field. Consequently, any Lorentz-covariant object, including propagators, self-energies, polarization tensors, vertex functions, and other scalar, vector, or tensor operators, must be decomposed into independent structures associated with these two sectors. Such a decomposition reflects the anisotropic dynamics induced by the background field and has important consequences for the properties of elementary excitations. In particular, this separation appears already at the classical level in the fermion--photon (quark--gluon) interaction, whose vertex acquires independent longitudinal and transverse components in the presence of the external field~\cite{Miransky:2015ava}. In the context of relativistic heavy-ion collisions, this anisotropy can lead to specific patterns for the decay of heavy resonances produced in the early stages of the reaction, providing a possible experimental handle to infer the strength of the magnetic field~\cite{Ayala:2025jpr}.

The already rich structure of the coupling between fermions and a neutral gauge-boson in the presence of a background magnetic field can be expected to become even richer when the neutral gauge-boson is replaced by a charged one such as the standard model $W^\pm$. For instance, it has been shown that the $W^\pm$ boson leptonic decay rate in a strong magnetic field is modified with respect to its vacuum value~\cite{Kurilin:2004qb}. The external field wave functions correspond to solutions of the equations of motion for the charged particles in the background field. In this calculation, these wave functions retain their polarization labels, but the calculation does not resolve the decay probability into separate channels defined by polarization relative to the external field.

Recall that the description of charged particles in external magnetic fields requires a proper treatment of Schwinger phases and magnetic translations. In Refs.~\cite{Carlomagno:2022arc,GomezDumm:2023owj} a consistent formalism in which charged mesons are expanded in Landau (Ritus) eigenfunctions, allows the meson polarization operator to be diagonalized in Landau-level space, leading to a gauge-invariant determination of the charged meson spectrum. In addition, the propagator properties of charged vector bosons in the presence of external electromagnetic fields have also received considerable attention over the years. Reference~\cite{Kuznetsov:2013sea} summarizes the findings for the proper-time description of the charged $W$-boson propagator in a constant magnetic field, providing gauge-covariant expressions that have served as the basis for numerous studies of electroweak processes in magnetized media, including neutrino propagation and weak interactions in astrophysical environments. More recently, Ref.~\cite{Iablokov:2020upc} generalized this formalism by deriving the propagator in arbitrary covariant gauges and clarifying its momentum-space and Landau-level representations. Building upon these developments, Ref.~\cite{Cancino:2026xpk} reformulated the charged-vector propagator within the Ritus eigenfunction method, constructing the complete set of spin-1 eigenfunctions in a magnetic field and establishing a LSZ reduction formalism for external charged vector bosons. These works have laid the theoretical foundations for a consistent description of charged $W$ bosons in external magnetic fields; however, the magnetic field dependence of the elementary $W q\bar q'$ interaction vertex itself has not yet been investigated. In this work, we undertake this calculation, finding the selection rules for the lowest order magnetic field modifications of the vertex coupling $W^\pm$ bosons with $\bar{q}q'$ quarks. The remainder of this paper is organized as follows. In Sec.~\ref{secII}, we introduce the process under consideration, review the structure of the interacting fields in the presence of a magnetic field, and establish the formalism required to construct the interaction vertex. In Sec.~\ref{secIII}, we focus on the process ($u\bar{d}\rightarrow W^+$) with both the quark and antiquark occupying the lowest Landau level (LLL) and their spins aligned with the magnetic field. This simplest configuration allows us to derive the first selection rules and elucidate their physical origin. In Sec.~\ref{secIV}, we extend the analysis to arbitrary fermion Landau levels. We first examine the tensor structure of the interaction, from which the spin and polarization selection rules naturally emerge, and then analyze the spatial contribution to determine the allowed Landau levels of the produced $W$-boson as a function of the Landau levels of the initial fermion and antifermion. In Sec.~\ref{secV}, we derive explicit expressions for the coefficients arising from the spatial overlap integral of three Hermite polynomials, with particular emphasis on the next-to-lowest Landau levels. These results are subsequently used to determine the corresponding selection rules and discuss their physical implications. Finally, Appendix~\ref{appA} summarizes the description of the $W$-boson polarization states in a magnetic field based on the formalism developed in Refs.~\cite{GomezDumm:2023owj,Iablokov:2020upc,Cancino:2026xpk}. As an independent validation of our theoretical framework, Appendix~\ref{appB} presents an alternative derivation of the spatial contribution governing the Landau-level selection rules and apply this approach to the next-to-lowest Landau-level examples discussed in the main text.

\section{Quark, antiquark and \texorpdfstring{$W$}{W}-boson wave functions in the Ritus representation}\label{secII}

We start our discussion by introducing the transition amplitude
(hereby referred to as the interaction vertex $\mathcal{V}$) that
constitutes the central focus of this analysis. The objective is
to derive the selection rules for the transition where a quark
(hereby labeled as $f$) and an antiquark (hereby labeled as
$\bar{f}$) annihilate to produce a $W$-boson in the final state.
This is given by
\begin{equation}
\label{vertx I}
    \mathcal{V} = V_{f_1,f_2} \int d^4x \Bra{W}\bar{\Psi}(x) \gamma^\mu \dfrac{(1-\gamma^5)}{2} \Psi(x) W_\mu(x) \Ket{f_1, \bar{f}_2} .
\end{equation}

where, to obtain the complete vertex structure, we insert the $V_{f_1,f_2}$ factor, corresponding to the Cabibbo-Kobayashi-Maskawa matrix element.

To facilitate the analytical developments and highlight their physical implications, we review in the following subsections the formalism describing charged particles in the presence of a background magnetic field.

\subsection{Fermion field in the Ritus representation}\label{secIIA}

To perform the vertex calculations, we adopt the formalism and notation established in Ref.~\cite{GomezDumm:2023owj}. Consequently, we provide a brief overview of the relevant theoretical results to be used. First, the fermion fields can be described as an expansion in the Ritus basis as
\begin{equation}\label{fermion field}
\begin{split}
    \psi_f(x) =& \sum\hspace{-2em}\int\limits_{\{\bar{q}_{E_f}\}} \sum_{a=1,2} \frac{1}{2E_f} \Big\{ b_f(\breve{q}, a) U_f(x, \bar{q}, a) \\
    &+ d_f(\breve{q}, a)^\dagger V_f(x, \bar{q}, a) \Big\},
\end{split}
\end{equation}
where the form of the sum-integral is determined by the gauge choice. Here, we take $\vec{B} = \left(0, 0, B \right)$ and work in the Landau gauge 2 (LG2) for which the electromagnetic field is given by
${\cal A}^\mu(x) = (0,0,B x^1,0)$. Therefore, the explicit form of the sum-integral is given by
\begin{equation}\label{def.sum-int}
    \begin{split}
        \sum\hspace{-2em}\int\limits_{\{\bar{q}_{E_f}\}} &\equiv  \sum_{k=0}^{\infty} \int \frac{dq^0}{2\pi} \frac{dq^2}{2\pi} \frac{dq^3}{2\pi} \delta(q^0-E_f) .
    \end{split}
\end{equation}
We also introduce the notation
\begin{equation}\label{def.delta(sum-int)}
    \begin{split}
         \delta_{\bar{q}\bar{q}'} &\equiv (2\pi)^4 \delta_{kk'} \delta(q^0 - q'^0) \delta(q^2 - q'^2) \delta(q^3 - q'^3).
    \end{split}
\end{equation}

Furthermore, within this framework, we define the momenta as $\bar{q} = (q^0, k, q^2, q^3)$ and $\breve{q} = (k, q^2, q^3)$.
Hereafter, $k$, $k'$ (and eventually $k_W$) refer to the Landau level occupied by the corresponding charged particle. In addition, we introduce the effective magnetic parameter
$B_f = |Q_f B|$ and the sign function $s = \text{sgn}(Q_f B)$, where $Q_f$ represents the charge of the respective particle (or antiparticle) under consideration. The momentum-space spinors, $U_f$ and $V_f$, introduced in Eq.~\eqref{fermion field}, are given by
\begin{align} \label{def. U and V}
U_f(x, \bar{q}, a) &= \mathbb{E}^{\mathcal{Q}_f}(x, \bar{q}) u_{Q_f}(k, q^3, a), \notag \\
V_f(x, \bar{q}, a) &= \tilde{\mathbb{E}}^{-\mathcal{Q}_f}(x, \bar{q}) v_{-Q_f}(k, q^3, a),
\end{align}
where $\mathbb{E}^{\mathcal{Q}}(x, \bar{q})$ and $\tilde{\mathbb{E}}^{-\mathcal{Q}}(x, \bar{q})$ denote the Ritus functions. Their explicit representations take the form
\begin{align}
\mathbb{E}^{\mathcal{Q}}(x, \bar{q}) &= \sum_{\lambda=\pm} \Gamma^\lambda \mathcal{F}_{\mathcal{Q}}(x, \bar{q}_\lambda), \nonumber \\
\tilde{\mathbb{E}}^{-\mathcal{Q}}(x, \bar{q}) &= \sum_{\lambda=\pm} \Gamma^\lambda \mathcal{F}_{-\mathcal{Q}}(x, \bar{q}_{-\lambda})^*,
\end{align}
In this framework, the spin projection operators are introduced as
$\Gamma^{\lambda} = (1 + \lambda S^3)/2$, where $S^3 =
i\gamma^1\gamma^2$ acts as the third component of the spin
operator within the spin-$1/2$ representation. Moreover, the
modified momentum variable is denoted by
$\bar{q}_\lambda = (q^0, n_f(k,s,\lambda), q^2,
q^3)$ with the
Landau-level-dependent indices shifted according to the relation
$n_f(k,s,\lambda) = k - (1 - \lambda s)/2$, where the index
$\lambda$ always takes the values $\pm$. Hereafter $n_f$, $n_f'$ (and eventually $n_W$) refer to the {\it oscillator index.} For particles with spin, $n_f$ depends
on the Landau-level index $k$, the sign of the electric charge, and the spin or polarization projection. Therefore, $n_f$ should not, in general, be identified with the physical Landau-level index.

Continuing with the properties of the Ritus functions, the spatial function $\mathcal{F}$ for the adopted gauge choice is expressed as

\begin{equation} \label{function F in LG2}
\mathcal{F}_{Q}(x, \bar{q}) = N_k e^{-i(q^0 x^0 - q^2 x^2 - q^3 x^3)} D_k(\rho_s),
\end{equation}
where the scaled coordinate is $\rho_s = \sqrt{2B_Q}(x^1 - s q^2 / B_Q)$ and the normalization constant reads as $N_k = (4\pi B_Q)^{1/4} / \sqrt{k!}$. The parabolic cylinder functions $D_k(x)$ present in the above expressions are defined in terms of the Hermite polynomials

\begin{equation}
D_k(x) = 2^{-k/2} e^{-x^2/4} H_k(x/\sqrt{2}),
\end{equation}
following the standard convention where $H_{-1}(x) = 0$.

\subsection{\texorpdfstring{$W$}{W}-boson field in the Ritus representation}\label{secIIB}
To describe the charged spin-1 boson, we once again adopt the notation established in Ref.~\cite{GomezDumm:2023owj}. Although the aforementioned work develops the formalism specifically for the $\rho$-meson, an identical framework can be directly applied to the $W$-boson, which is the object of interest here. The boson field is given by the following expansion
\begin{equation}\label{boson field}
\begin{split}
    W^{\mathcal{Q},\mu}(x) =& \ \sum\hspace{-2em}\int\limits_{\{\breve{q}_{E_W}\}} \sum_c \frac{1}{2E_W} \Big[ a_W^{\mathcal{Q}}(\breve{q}, c) W^\mu_{\mathcal{Q}}(x, \bar{q}, c) \\
    &+ a_W^{-\mathcal{Q}}(\breve{q}, c)^\dagger W^{\mu}_{-\mathcal{Q}}(x, \bar{q}, c)^* \Big].
\end{split}
\end{equation}
The definitions of the momenta $\bar{q}$ and $\breve{q}$ follow those introduced for the fermion case, along with the effective magnetic parameter $B_W = |Q_W B|$ and the sign function $s = \text{sgn}(Q_W B)$. It is crucial to emphasize that for a boson-vector field interacting with a magnetic background, the sum over the integer index $k$ (associated with the Landau levels) starts at $k = -1$.

The wave functions $W^\mu_{\mathcal{Q}}(x, \bar{q}, c)$ in the above expansion are factored as follows
\begin{equation}
W^\mu_{\mathcal{Q}}(x, \bar{q}, c) = \mathbb{R}^{\mathcal{Q},\mu\nu}(x, \bar{q}) \epsilon_{\mathcal{Q},\nu}(k, q^3, c),
\end{equation}
where we have separated the spatial and momentum dependence, contained within the tensor $\mathbb{R}^{\mathcal{Q},\mu\nu}$, from the polarization vector $\epsilon_{\mathcal{Q},\nu}(k, q^3, c)$, with $c$ being the index denoting the polarization state. Due to the orthogonality relation of these vectors, it is noteworthy that for the lowest Landau level ($k = -1$), there is only one possible polarization. Hence, the index $c$ exclusively takes the value $c = 1$; see Appendix~\ref{appA}.

The spatial and tensorial part, $\mathbb{R}^{\mathcal{Q},\mu\nu}$, is expressed through the sum

\begin{equation}
\mathbb{R}^{\mathcal{Q},\mu\nu}(x, \bar{q}) = \sum_{\lambda=-1,0,1} \mathcal{F}_{\mathcal{Q}}(x, \bar{q}_\lambda) \Upsilon^{\mu\nu}_\lambda,
\end{equation}
where the modified momentum for the boson is defined as $\bar{q}_\lambda = (q^0, n_{W}, q^{2}, q^3)$, where $n_{W} \equiv k - s\lambda$. Notice that this definition differs from the one used for the fermions. Finally, the tensors $\Upsilon^{\mu\nu}_\lambda$, which project the field components with respect to the background magnetic field, are explicitly given by
\begin{equation}
\Upsilon^{\mu\nu}_0 = g^{\mu\nu}_\parallel, \quad \Upsilon^{\mu\nu}_{\pm 1} = \frac{1}{2}(g^{\mu\nu}_\perp \mp S^{\mu\nu}_3),
\label{upsilonmunu}
\end{equation}
with
\begin{equation}
S_3^{\mu\nu} = i(\delta^\mu_1 \delta^\nu_2 - \delta^\mu_2 \delta^\nu_1).
\label{S3munu}
\end{equation}

For the remaining definitions and mathematical structures required to complete the theoretical formulation of each field in this notation, as well as for a comprehensive discussion regarding its relation and modifications compared to other formalisms in the literature, we refer the reader to Ref.~\cite{GomezDumm:2023owj}.

\subsection{Formulation of the quark-antiquark-\texorpdfstring{$W$}{W} vertex}
Using the structures defined in the preceding sections and starting from Eq. \eqref{vertx I},
we proceed to write the vertex describing the annihilation of the two initial fermions leading to the subsequent creation of a $W$-boson in the final state.
For this purpose, we employ standard quantum field theory techniques that include the boson and fermion commutation relations. Notice that since the creation and
annihilation operators describe physical asymptotic states on the mass shell, the spatial and momentum conserving delta functions in these relations correspond exactly
to the 3-dimensional on-shell projection of the full orthogonality relation previously defined in Eq.~\eqref{def.delta(sum-int)}. For practical purposes, it is useful to define this 3-dimensional projection as
\begin{equation}
    \delta_{\breve{q}\breve{q}'} \equiv (2\pi)^3 \delta_{kk'}\delta(q^2-q'^2)\delta(q^3-q'^3),
\end{equation}
such that $\delta_{\bar{q}\bar{q}'} = 2\pi\delta(q^0-q'^0)\delta_{\breve{q}\breve{q}'}$. The commutation and anticommutation relations are then given, respectively, by

\begin{align}
[a_{W}^{\mathcal{Q}}(\breve{q}, c), a_{W}^{\pm \mathcal{Q}}(\breve{q}', c')] =& [a_{W}^{\mathcal{Q}}(\breve{q}, c)^{\dagger}, a_{W}^{\pm \mathcal{Q}}(\breve{q}', c')^{\dagger}] \notag \\
=& [a_{W}^{\mathcal{Q}}(\breve{q}, c), a_{W}^{-\mathcal{Q}}(\breve{q}', c')^{\dagger}] = 0, \notag \\
[a_{W}^{\mathcal{Q}}(\breve{q}, c), a_{W}^{\mathcal{Q}}(\breve{q}', c')^{\dagger}] =& [a_{W}^{-\mathcal{Q}}(\breve{q}, c), a_{W}^{-\mathcal{Q}}(\breve{q}', c')^{\dagger}] \notag \\
=& 2E_{W}\delta_{cc'}\delta_{\breve{q}\breve{q}'};
\end{align}
and
\begin{align}
\{b_{f}(\breve{q}, a), b_{f}(\breve{q}', a')\} =& \{d_{f}(\breve{q}, a), d_{f}(\breve{q}', a')\} = 0, \notag \\
\{b_{f}(\breve{q}, a), d_{f}(\breve{q}', a')\} =& \{b_{f}(\breve{q}, a), d_{f}(\breve{q}', a')^{\dagger}\} = 0, \notag \\
\{b_{f}(\breve{q}, a), b_{f}(\breve{q}', a')^{\dagger}\} =& \{d_{f}(\breve{q}, a), d_{f}(\breve{q}', a')^{\dagger}\} \notag \\
=& 2E_{f}\delta_{aa'}\delta_{\breve{q}\breve{q}'}.
\end{align}
Using these relations and substituting the field expansions provided in Eqs.~\eqref{fermion field} and~\eqref{boson field}, the vertex of interest can be explicitly expressed as
\begin{equation}\label{vertx II}
    \mathcal V = V_{f_1,f_2} \int d^4 x \bar{V}_{\bar{f}_2} \gamma^\mu \dfrac{(1-\gamma^5)}{2} U_{f_1} W^*_{Q_W , \mu}.
\end{equation}
Thus, employing the definitions detailed in Secs.~\ref{secIIA} and \ref{secIIB}, we can establish the general form of the vertex by separating it into a tensor part and a spatial part associated with the $d^4x$ integration. The vertex is thus given by
\begin{widetext}
    \begin{equation}\label{vertx III}
    \begin{split}
        \mathcal V =& V_{f_1,f_2} \sum_{\lambda_1,\lambda_2 = \pm} \sum_{\lambda_W = -1,0,1} \left[\bar{v}_{-Q_2} \Gamma^{\lambda_2} \gamma^\mu \dfrac{(1-\gamma^5)}{2} \Gamma^{\lambda_1} u_{Q_1}\right]\Upsilon_{\lambda_W, \mu\nu}^* \epsilon^{\nu *}_{Q_W} \left[\int d^4x\mathcal{F}_{-Q_2}\mathcal{F}_{Q_1}\mathcal{F}^*_{Q_W}\right].
        \end{split}
    \end{equation}
\end{widetext}
 This element dictates the strength of the flavor-changing transition between the quarks, mediated by the charged weak interaction of the $W$ boson.
\section{Quark--antiquark--\texorpdfstring{$W$}{W}-boson vertex in the lowest Landau level}\label{secIII}

To extract physical information from the vertex, we investigate a particular case of interest: the $u\bar{d} \rightarrow W^+$ process, imposing that both the initial quark and antiquark occupy the LLL and consequently have their spins aligned with the background magnetic field. Thus, employing Eq.~\eqref{vertx III}, the vertex for this particular case is expressed as

\begin{equation}\label{vertx udW}
\begin{split}
            \mathcal V =& \ V_{u,d}\sum_{\lambda_{\bar d},\lambda_{u} = \pm} \sum_{\lambda_W = -1,0,1} \left[\bar{v}_{-Q_d} \Gamma^{\lambda_{\bar d}} \gamma^\mu \dfrac{(1-\gamma^5)}{2} \Gamma^{\lambda_u} u_{Q_u}\right] \\
        &\times \Upsilon_{\lambda_W, \mu\nu}^* \epsilon^{\nu *}_{Q_W} \left[\int d^4x\mathcal{F}_{-Q_d}\mathcal{F}_{Q_u}\mathcal{F}^*_{Q_{W}}\right].
\end{split}
\end{equation}

As previously stated in Sec.~\ref{secII}, we can divide our analysis into two distinct parts. First, the tensor structure dictates how the initial spin configurations determine the polarization state of the generated boson. Second, the spatial part relates the chosen Landau levels of the initial fermions to the constraints imposed on the $W^+$ boson.

\subsection{Tensor structure for the quark--antiquark--\texorpdfstring{$W$}{W} vertex}\label{secIIIA}

To verify our framework, we derive the selection rules emerging from the tensor structure, which encapsulates the entire dependence on the quark spin projections and the boson polarization. We show that the selection rules governing the spin projections and polarization are contained within the first term of Eq.~\eqref{vertx udW}, which we denote as
\begin{eqnarray}
        \tau^{\lambda_u\lambda_{\bar d},\lambda_W} &=& \left[ \bar{v}_{-Q_{\bar d}} \Gamma^{\lambda_{\bar d}}\left(\gamma^\mu_{||} + \gamma^\mu_\perp\right)\Gamma^{\lambda_u} \dfrac{(1-\gamma^5)}{2}u_{Q_u}\right]\nonumber\\
        &\times& \Upsilon_{\lambda_W, \mu\nu}^* \epsilon^{\nu *}_{Q_W}.
\end{eqnarray}

The first step is to expand the spin projectors. This produces the initial Kronecker deltas of this structure. This procedure is carried out as follows
\begin{widetext}
    \begin{equation}\label{tensor udW}
    \begin{split}
        \tau^{\lambda_u\lambda_{\bar d},\lambda_W} =& \left[ \bar{v}_{-Q_d} \Gamma^{\lambda_{\bar d}}\left(\gamma^\mu_{||} + \gamma^\mu_\perp\right)\Gamma^{\lambda_u} \dfrac{(1-\gamma^5)}{2}u_{Q_u}\right]\Upsilon_{\lambda_W, \mu\nu}^* \epsilon^{\nu *}_{Q_W} \\
        =& \left[ \bar{v}_{-Q_d} \left(\gamma^\mu_{||} \Gamma^{\lambda_{\bar d}}\Gamma^{\lambda_u} + \gamma^\mu_\perp  \Gamma^{-\lambda_{\bar d}}\Gamma^{\lambda_u}\right) \dfrac{(1-\gamma^5)}{2}u_{Q_u}\right]\Upsilon_{\lambda_W, \mu\nu}^* \epsilon^{\nu *}_{Q_W}\\
         =& \left[ \bar{v}_{-Q_d} \left(\gamma^\mu_{||} \Gamma^{\lambda_u} \delta_{\lambda_u,\lambda_{\bar d}} + \gamma^\mu_\perp \Gamma^{\lambda_u}\delta_{\lambda_u,-\lambda_{\bar d}}\right) \dfrac{(1-\gamma^5)}{2}u_{Q_u}\right]\Upsilon_{\lambda_W, \mu\nu}^* \epsilon^{\nu *}_{Q_W} \\
    \end{split}
\end{equation}
\end{widetext}
By introducing the tensor $\Upsilon^*_{\lambda_W, \mu\nu}$ into the parentheses, we perform a contraction over the index $\mu$, which is restricted to specific components due to the presence of the $\gamma_{\parallel}$ and $\gamma_\perp$ matrices. Consequently, we must employ the definition of the tensors given in Eqs.~(\ref{upsilonmunu}) and~(\ref{S3munu}).
For each term, we have the following relations
\begin{equation}
    \gamma^\mu_{||} \Upsilon_{\lambda_W, \mu\nu}^* = \gamma^\mu_{||}  g_{\mu\nu} \delta_{0,\lambda_W} = \gamma_{||,\nu}  \delta_{0,\lambda_W}
\end{equation}
and
\begin{equation}
    \begin{split}
        \gamma_{\perp}^\mu \Upsilon_{\lambda_W, \mu\nu}^* =&\gamma_{\perp}^\mu \Upsilon_{-\lambda_W, \mu\nu} (1-\delta_{0,\lambda_W})\\
        =& \gamma_{\perp,\nu} \Gamma^{-\lambda_W} (1-\delta_{0,\lambda_W}) ,
    \end{split}
\end{equation}
where we used the property $\gamma_{\perp}^\mu S_{3,\mu\nu} = -\gamma_{\perp,\nu}S^3$. Furthermore, it is justifiable to use the term $\Gamma^{-\lambda_W}$, as it is accompanied by a delta that restricts the selection $\lambda_W = \pm 1$. Consequently, in this term, we will never encounter the $\lambda_W=0$ case, which would invalidate the definition of the projector $\Gamma^\lambda$.

Returning to the results in Eq.~\eqref{tensor udW}, we obtain
\begin{widetext}
    \begin{equation}\label{tensor udW F}
    \begin{split}
        \tau^{\lambda_u,\lambda_{\bar d}, \lambda_W}=& \left[ \bar{v}_{-Q_d} \Bigg(\gamma_{||,\nu}   \Gamma^{\lambda_u} \delta_{0,\lambda_W}\delta_{\lambda_u,\lambda_{\bar d}} + \gamma_{\perp,\nu} \Gamma^{-\lambda_W}\Gamma^{\lambda_u} (1-\delta_{0,\lambda_W}) \delta_{\lambda_u, -\lambda_{\bar d}} \Bigg) \dfrac{(1-\gamma^5)}{2}u_{Q_u}\right] \epsilon^{\nu *}_{Q_W}\\
              =& \left[  \bar{v}_{-Q_{d}} \Bigg(  \delta_{\lambda_W,0} \ \delta_{\lambda_u,\lambda_{\bar d}}\ \gamma_{||,\nu} + \delta_{-\lambda_W ,\lambda_u}
        \delta_{\lambda_u, -\lambda_{\bar d}} \ \gamma_{\perp,\nu} \Bigg)  \Gamma^{\lambda_u} \, \dfrac{(1-\gamma^5)}{2} u_{Q_u} \right
        ] \ \epsilon^{\nu *}_{Q_W} .
    \end{split}
\end{equation}
\end{widetext}

\subsection{Space-time structure with the quark-antiquark in the LLL}\label{secIIIB}
To derive the selection rules corresponding to the allowed Landau levels for the production of the $W$-boson in the particular case of interest, we shall focus our analysis on the spatial part of Eq.~\eqref{vertx udW}, which is given by the integral
\begin{equation}
    I = \int d^4x \mathcal F_{-Q_d}(x, \bar p_{2,-\lambda_{\bar d}}) \mathcal F_{Q_u}(x, \bar p_{2,\lambda_u})\mathcal F^*_{Q_W}(x, \bar q_{\lambda_W}).
\end{equation}

Before applying the definitions introduced in Eq.~\eqref{function F in LG2}, into the above equation, it is useful to note that under a gauge transformation
$\Lambda(x)$ the functions $\mathcal{F}_Q(x,\bar{q})$ transform as
\begin{equation}\label{F transformation}
\mathcal{F}_Q(x,\bar{q}) \rightarrow \tilde{\mathcal{F}}_Q(x,\bar{q})
= e^{-iQ\Lambda(x)}\mathcal{F}_Q(x,\bar{q}).
\end{equation}
Consequently, since electric charge conservation dictates that $Q_u - Q_d = Q_W$, the gauge phase factors exactly cancel.
Thus, the spatial part of the vertex under consideration is manifestly gauge invariant. The framework developed thus far could therefore have been formulated without
fixing a gauge from the outset. In that case, any of the representations introduced in Ref.~\cite{GomezDumm:2023owj} could be used to perform our analysis. Nevertheless, to remain consistent with the conventions established in Sec.~\ref{secII}, we continue working in the LG2.
Substituting the corresponding expressions for these functions into the integral for the specific case considered here, we obtain
\begin{widetext}
    \begin{equation}
    \begin{split}
        I =& \int d^4x \left[N_{n_{\bar d}} \ e^{-i(p_{\bar d}^0x^0-p_{\bar d}^2x^2-p_{\bar d}^3x^3)} D_{n_{\bar d}}\left(\sqrt{2B_{\bar d}}\left(x^1-s_{\bar d}\dfrac{p_{\bar d}^2}{B_{\bar d}}\right) \right)\right] \Bigg[N_{n_u} e^{-i(p_u^0x^0-p_u^2x^2-p_u^3x^3)} \\
        &\times D_{n_u}\left(\sqrt{2B_u}\left(x^1-s_u\dfrac{p_u^2}{B_u}\right) \right)\Bigg] \left[N_{n_W} e^{i(q_W^0x^0-q_W^2x^2-q_W^3x^3)} D_{n_W}\left(\sqrt{2B_W}\left(x^1-s_W\dfrac{q_W^2}{B_W}\right) \right)\right].
    \end{split}
\end{equation}
\end{widetext}

At this stage, we can initially evaluate the free part described by the exponentials and factor out the constants $N_{n_i}$, yielding
\begin{equation}
    \begin{split}
        I =& N_{n_u}N_{n_{\bar d}}N_{n_W} \int dx^0dx^2dx^3 \Big[e^{-i(p_u^0+p_{\bar d}^0-q_W^0)}e^{i(p_u^2+p_{\bar d}^2-q_W^2)} \\
        &\times e^{i(p_u^3+p_{\bar d}^3-q_W^3)} \Big]\int dx^1 \left[D_{n_u}(\rho_{s_u})D_{n_{\bar d}}(\rho_{s_{\bar d}})D_{n_W}(\rho_{s_W})\right] \\
        =& N_{n_u}N_{n_{\bar d}}N_{n_W} (2\pi)^3  \delta(p_u^0+p_{\bar d}^0-q_W^0)\delta(p_u^2+p_{\bar d}^2-q_W^2) \\
        &\times\delta(p_u^3+p_{\bar d}^3-q_W^3) \int dx^1 \Big[D_{n_u}(\rho_{s_u})D_{n_{\bar d}}(\rho_{s_{\bar d}})\\
        &\times D_{n_W}(\rho_{s_W})\Big].
    \end{split}
\end{equation}
To simplify the notation, we define the following variables
\begin{align}
    x_{c_u} = \dfrac{s_u p_u^2}{B_u} \ ; \ x_{c_{\bar d}} = \dfrac{s_{\bar d}p_{\bar d}^2}{B_{\bar d}} \ ; \ x_{c_W} = \dfrac{s_Wq_W^2}{B_W},
\end{align}
together with
\begin{equation}
    \mathcal N (n_u,n_{\bar d},n_W)= 2^{\left(-\frac{n_u}{2}\right)} \ 2^{\left(-\frac{n_{\bar d}}{2}\right)} \ 2^{\left(-\frac{n_W}{2}\right)}.
\end{equation}
Thus, focusing solely on the remaining integral and employing the definitions of $D_n(\rho_{s_n})$, the integration over the variable $x^1$ yields
\begin{eqnarray}\label{int dx1 udW}
        I_D&=& \ \mathcal N (n_u,n_{\bar d},n_W)\nonumber\\
        &\times&\int dx^1 \exp\left[-\dfrac{1}{2} \sum_{i = u,\bar d,W} B_i(x^1-x_{c_i})^2 \right]\nonumber \\
        &\times& H_{n_u}(\rho_u/\sqrt{2}) H_{n_{\bar d}}(\rho_{\bar d}/\sqrt{2}) H_{n_W}(\rho_W/\sqrt{2}).
\end{eqnarray}
However, since we are investigating a specific case where the initial particles are confined to the LLL, we can impose the conditions $n_u = 0$ and $n_{\bar d} = 0$ on the functions.
Using $H_0(x)=1$, the integral in Eq.~\eqref{int dx1 udW} becomes
    \begin{equation}
    \begin{split}
        I_D
        =& \ 2^{-n_W/2} \int dx^1 \exp\left[-\dfrac{1}{2} \sum_{i = u,\bar d,W} B_i(x^1-x_{c_i})^2 \right] \nonumber \\ & H_{n_W}\Big( \sqrt{B_W}(x^1-x_{c_W})\Big).
    \end{split}
\end{equation}

Now, manipulating the argument of the exponential and using the conservation of momentum along the $x^2$ direction, we note that
\begin{equation}
    \begin{split}
        & \sum_{i = u,\bar d,W} B_i(x^1-x_{c_i})^2  = 2B_W(x^1-x_{c_W})^2 + \Theta\\
    \end{split}
\end{equation}
where
\begin{eqnarray}
\Theta = B_u (x_{c_u})^2 + B_{\bar d} \ (x_{c_{\bar d}})^2 - B_W (x_{c_W})^2
\label{theta}
\end{eqnarray}
Substituting this back into the integral, we observe that the dependence on the variable $x^1$ is simplified, yielding
\begin{equation}\label{I_D simplify}
    \begin{split}
        I_D =& \ 2^{-\frac{n_W}{2}} e^{-\frac{\Theta}{2}} \int dx^1 e^{-B_W(x^1-x_{c_W})^2} \\
        &\times H_{n_W}\Big( \sqrt{B_W}(x^1-x_{c_W})\Big).
    \end{split}
\end{equation}
Applying the change of variables $y = \sqrt{B_W} (x^1 - x_{c_W})$, the integral \eqref{I_D simplify} becomes
\begin{equation}\label{I_D change variable}
    I_D = \dfrac{2^{-\frac{n_W}{2}} e^{-\frac{\Theta}{2}}}{\sqrt{B_W}} \int dy \ e^{-y^2} H_{n_W}(y).
\end{equation}

The resulting integral in Eq~\eqref{I_D change variable} can be evaluated simply by recalling that $H_{0} = 1$. Consequently, we can employ the orthogonality relation of the Hermite polynomials, which ensures that the product of two polynomials of different degrees yields

\begin{equation}\label{orth. hermite}
    \int dx H_m(x)H_n(x) e^{-x^2} = \sqrt{\pi} 2^n n! \delta_{nm}.
\end{equation}
Therefore, the result obtained for the integral under the established conditions is
\begin{equation}
    \begin{split}
        I_D =& \ \dfrac{2^{-\frac{n_W}{2}} e^{-\frac{\Theta}{2}}}{\sqrt{B_W}} \int dy \ e^{-y^2} H_0(y) H_{n_W}(y) \\
        =& \  \sqrt{\dfrac{\pi}{B_W}} 2^{-\frac{n_W}{2}} e^{-\frac{\Theta}{2}} \delta_{n_W,0}.
    \end{split}
\end{equation}
By applying the resulting delta function to the constants previously factored outside the integral, we can conclude that the spatial integral is

\begin{widetext}
    \begin{equation}\label{result spatial udW}
    \begin{split}
        I
        =& \ (4\pi B_u)^{1/4}(4\pi B_{\bar d})^{1/4}(4\pi B_W)^{1/4} \ \delta_{023}(p_u+p_{\bar d} - q_W) \ e^{-\frac{\Theta}{2}} \ \sqrt{\dfrac{\pi}{B_W}} \ \delta_{n_W,0},
    \end{split}
    \end{equation}
where
\begin{eqnarray}
\delta_{023}(p+p' - q) = (2\pi)^3  \delta(p^0+{p'}^0-q^0)\delta(p^2+{p'}^2-q^2) \delta(p^3+{p'}^3-q^3)
\label{deltas}
\end{eqnarray}
\end{widetext}
Therefore, by imposing the initial condition whereby the quark and the antiquark both occupy the LLL, the spatial integral in the vertex calculation dictates that the generated $W$-boson must be in a state with $n_W=0$. As we show in Sec.~\ref{secIIIC}, this requirement also implies that the $W$-boson must occupy the LLL.

\subsection{Selection rules}\label{secIIIC}
By imposing that the initial quarks in the $u\bar{d} \rightarrow W^+$ process occupy the LLL, we obtained a selection rule dictating that the generated $W^+$ boson must be in a state with $n_W=0$. Here we show that such selection implies that the $W$-boson must also occupy the LLL (i.e. $k_W=-1$).
According to Refs.~\cite{GomezDumm:2023owj,Iablokov:2020upc,Cancino:2026xpk}, in that state, the only permissible polarization for a vector boson is given by
$c = 1$ (see Appendix~\ref{appA}).
To verify whether the formalism developed here reflects the theoretical foundations and physical expectations, we first analyze the projection indices $\lambda$ for the initial particles. In this section, we impose that the $u$ and $\bar{d}$ quarks have their spins aligned with the background magnetic field. This assumption is physically consistent since both particles have positive charges and, in the LLL, the only permitted spin state is the one aligned with the field.
Therefore, assuming $B > 0$ and that both particles are in the $k = 0$ (LLL) state, we can use that for fermions
\begin{equation}
    n_f(k,s,\lambda) = k- \Big(\dfrac{1-s\lambda}{2}\Big),
\end{equation}
and the corresponding expression for antifermions
\begin{equation}
    n_{\bar f}(k,s,-\lambda) = k - \Big(\dfrac{1+s\lambda}{2}\Big),
\end{equation}
to determine their respective projections that ensure alignment with the field.

First, for the $u$-quark, we have $s_u = \text{sgn}(Q_u B) = +1$. Imposing $k_u = 0$, we obtain

\begin{eqnarray}
    n_u(0,+1,\lambda_u) = - \Big(\dfrac{1-\lambda_u}{2}\Big) = 0 \Rightarrow \lambda_u = +1 .
\end{eqnarray}
On the other hand, for the $\bar{d}$ antiquark, the sign function yields $s_{\bar{d}} = \text{sgn}(Q_{\bar{d}} B) = +1$, and using $k_{\bar{d}} = 0$, we find that
\begin{eqnarray}
    n_{\bar d}(0,+1,\lambda_{\bar d}) = - \Big(\dfrac{1+\lambda_{\bar d}}{2}\Big) = 0 \Rightarrow \lambda_{\bar d}= -1 .
\end{eqnarray}
Thus, to project both spins along the direction of the magnetic field and determine which boson polarization state is selected under these conditions, we shall adopt $\lambda_u = 1$ and $\lambda_{\bar{d}} = -1$.

Substituting these results into the tensor structure developed in Eq. \eqref{tensor udW F}, we obtain
\begin{widetext}
    \begin{equation}\label{tensor udW II}
    \begin{split}
        \tau^{+1,-1, \lambda_W}=&  \bar{v}_{-Q_{d}} \Bigg( \delta_{\lambda_W,0} \ \cancelto{0}{\delta_{+1,-1}} \ \ \ \gamma_{||,\nu} +
          \delta_{-\lambda_W ,+1} \ \ \cancelto{1}{\delta_{+1, -(-1)}} \ \ \ \gamma_{\perp,\nu} \Bigg) \Gamma^{+} \dfrac{(1-\gamma^5)}{2}u_{Q_u}  \ \epsilon^{\nu *}_{Q_W} \\
        =&  \delta_{-\lambda_W ,+1} \left[  \bar{v}_{-Q_{d}} \ \gamma_{\perp,\nu} \ \Gamma^{+} \ \dfrac{(1-\gamma^5)}{2} \ u_{Q_u} \right] \epsilon^{\nu *}_{Q_W} .
    \end{split}
\end{equation}
\end{widetext}
The non-vanishing component selected due to the opposite signs of $\lambda$ corresponds to the perpendicular part, which in turn selects the index $\lambda_W = -1$ for the boson. This result, together with Eq.~\eqref{result spatial udW}, shows that the $W^+$ boson must be produced in the LLL. The physical expectation dictated by the dispersion relation
\begin{eqnarray}\label{LLL W+}
    n_W &=& k_W - s\lambda_W = 0 \nonumber\\
    &\rightarrow& (-1) - \lambda_W = 0 \Rightarrow\lambda_W = -1,
\end{eqnarray}
is strictly confirmed by the constructed tensor formalism. Consequently, obtaining $\lambda_W = -1$ corroborates that the boson occupies the LLL assuming the single allowed polarization state, as detailed in Appendix~\ref{appA}.

In summary, in this section, we have analyzed the particular case of the $u\bar{d} \rightarrow W^+$ vertex. For this scenario, we imposed the initial conditions whereby the $u$ and $\bar{d}$ quarks occupy the LLL with their spins aligned with the magnetic field. As a result of our theoretical framework, we demonstrated that the generated $W^+$ boson must necessarily occupy the LLL and possess a spin projection aligned with the background field (see Eq.(\ref{wspin})). This conclusion perfectly corroborates the physical expectations, given that $c=1$ is the only permitted polarization state for
the boson in its lowest Landau level.

\section{General structure of the lowest order quark--antiquark--\texorpdfstring{$W$}{W}-boson vertex in the presence of a magnetic field}\label{secIV}

In this section, we develop a general framework applicable to a process of the form $f_1 \bar{f}_2 \rightarrow W^{\pm}$ for arbitrary Landau levels. To this end, we proceed directly from the general vertex Eq.~\eqref{vertx III} and, analogous to the previous procedure, separate our analysis into two components: the tensor structure and the spatial integration. As a result,
we determine not only the allowed polarization states for the boson but also the allowed oscillator indices accessible to the generated $W^{\pm}$, given the constraints
imposed by the initial conditions of the fermion and antifermion indices, $n_1$ and $n_2$, respectively.

\subsection{Tensor structure}\label{secIVA}
Regarding the tensor structure, which establishes the selection rules arising from the chosen spin projections of fermion 1 and antifermion 2, the solution is strictly identical to the one developed in Sec.~\ref{secIIIA}. This equivalence is due to the fact that, in the derivation of this component, no approximations or simplifications restricted to the LLL were required.

Therefore, we simply rewrite the result obtained in Eq.~\eqref{tensor udW F} by performing the substitutions $u \rightarrow f_1$ and $\bar{d} \rightarrow \bar{f}_2$. Thus, the generalized expression for the tensor part is given by
\begin{equation}\label{tensor udW F2}
    \begin{split}
        & \tau^{\lambda_1,\lambda_2, \lambda_W}= \bar{v}_{-Q_2} \Bigg( \delta_{0,\lambda_W}\delta_{\lambda_1,\lambda_2} \gamma_{||,\nu}  +   \\
        & \qquad \qquad  \delta_{-\lambda_W ,\lambda_1} \delta_{\lambda_1, -\lambda_2} \gamma_{\perp,\nu} \Bigg) \Gamma^{\lambda_1} \dfrac{(1-\gamma^5)}{2}u_{Q_1} \ \epsilon^{\nu *}_{Q_W} . \qquad
    \end{split}
\end{equation}

\subsection{Space-time structure}\label{secIVB}
Having established the tensor structure, we now proceed to derive the general expression for the spatial part of the transition amplitude.
This allows us to determine the selection rules governing the production of a charged $W$-boson from fermions occupying arbitrary Landau levels.
As demonstrated below, these rules imply that the Landau level of the $W$-boson cannot be arbitrary but is restricted to a finite interval determined by the combined
Landau levels of the initial fermions. We emphasize that, as in the particular case discussed in Sec.~\ref{secIII}, the integral is gauge invariant owing to the transformation
properties of the spatial functions given in Eq.~\eqref{F transformation}. Adopting LG2 once again, the analysis of the spatial contribution in its most general form begins with the integral
\begin{equation}
    \begin{split}
        I =& \ N_{n_1}N_{n_2}N_{n_W} \delta_{023}(p_1+p_2-q_W) \\
        & \times  \int dx^1 D_{n_1}D_{n_2}D_{n_W}.
    \end{split}
\end{equation}
where $\delta_{023}$ has been defined in Eq.(\ref{deltas}). We focus our attention on the only remaining integral, which we denote as
\begin{eqnarray}
        I_D &=& \int dx^1 \ D_{n_1}(\rho_1)D_{n_2}(\rho_2)D_{n_W}(\rho_W)  \nonumber\\
        &=& \mathcal{N}(n_1,n_2,n_W)\nonumber\\
        &\times&\int dx^1 \exp \left[-\dfrac{1}{2} \sum_{i=1,2,W} B_i(x^1 - x_{c_i})^2\right] \nonumber\\
        &\times& H_{n_1}(\rho_1/\sqrt2)H_{n_2}(\rho_2/\sqrt2)H_{n_W}(\rho_W/\sqrt2),
    \label{int I_D geral}
\end{eqnarray}
where the explicit form of $\rho_i$ was introduced in Sec.~\ref{secIIIA}. Additionally, Eq.~\eqref{int I_D geral} introduces the definition $\mathcal{N}(n_1,n_2,n_W) = 2^{-\frac{n_1}{2}}2^{-\frac{n_2}{2}}2^{-\frac{n_W}{2}}$, while the orbit center coordinates, $x_{c_i}$, remain identical to those defined in the previous section.

We now analyze the sum in the argument of the exponential, given that we are dealing with $s_i = \pm$. After a bit of straightforward algebra, it can be shown that
\begin{equation}\label{sum exp var. shift}
        \sum_{i=1,2,W} B_i(x^1 - x_{c_i})^2
        = 2B_W \left[x^1 - X_c \right]^2 + \Delta,
\end{equation}
where the term $\Delta$ represents a contribution independent of the integration variable similar to the one appearing in Eq. \eqref{theta}.

The term $X_c$, in turn, emerges from the argument of the quadratic term involving the $x^1$ variable. This factor acts as a spatial coordinate shift, analogous to the orbit center displacements observed within the arguments of the Hermite polynomials. We define it as
\begin{equation}
    \begin{split}
        X_c = \dfrac{(B_1x_{c_1}+B_2x_{c_2}+B_Wx_{c_W})}{(B_1+B_2+B_W)}.
    \end{split}
\end{equation}

The entire procedure is carried out in a manner analogous to the previous section. However, here we make the formalism explicit for an arbitrary choice of signs. By applying this expansion to the integral, Eq.~\eqref{int I_D geral} takes the form
\begin{equation}
\begin{split}
        I_D =& \ \mathcal N e^{-\frac{\Delta}{2}} \int dx^1 e^{-B_W(x^1 -X_c)^2} H_{n_1}\left(\sqrt{B_1}(x^1-x_{c_1})\right) \\
        &\times H_{n_2}\left(\sqrt{B_2}(x^1-x_{c_2})\right)H_{n_W}\left(\sqrt{B_W}(x^1-x_{c_W})\right).
\end{split}
\end{equation}

The crucial aspect at this stage is that generating a $W^\pm$ boson strictly requires $s_1 = s_2 = s_W = \pm 1$. Given this condition imposed on the charges, the signs will be identical for all particles, allowing this factor to be absorbed into the notation used to define the shift of the orbit center. This means in particular that $\Theta = \Delta$. A direct consequence of this argument is that momentum conservation along the $x^2$ direction will always be satisfied. Therefore, we can express the shift $X_c$ simply as
\begin{equation}
    \begin{split}
        X_c =& \ \dfrac{B_1 x_{c_1}+B_2 x_{c_2}+B_W x_{c_W}}{2B_W} = \dfrac{p_1^2+p_2^2+B_Wx_{c_W}}{2B_W} \\
        =& \ \dfrac{p^2_W+B_Wx_{c_W}}{2B_W} = \dfrac{B_Wx_{c_W}+B_Wx_{c_W}}{2B_W} = x_{c_W}.
    \end{split}
\end{equation}
Thus, the argument of the exponential becomes identical to that of the boson's Hermite polynomial, which significantly facilitates the visualization of the previous case solution in a more general form. Following the same procedure, we apply the following change of variables

\begin{equation}
\begin{split}
        &y = \sqrt{B_W}(x^1-x_{c_W}) \ \Rightarrow dy = \sqrt{B_W} dx^1 \ ; \\
        &x^1 = \dfrac{y}{\sqrt{B_W}} + x_{c_W},
\end{split}
\end{equation}
which transforms the integral into
\begin{equation}
    \begin{split}
            I_D =& \ \mathcal{N}(n_1,n_2,n_W) e^{-\frac{\Theta}{2}}\int \dfrac{dy}{\sqrt{B_W}} e^{-y^2}\\
            &\times H_{n_1}\Bigg(\sqrt{B_1}\left(\dfrac{y}{\sqrt{B_W}} +(x_{c_W}-x_{c_1}) \right)\Bigg)\\
            &\times H_{n_2}\Bigg(\sqrt{B_2}\left(\dfrac{y}{\sqrt{B_W}} +(x_{c_W}-x_{c_2}) \right)\Bigg)H_{n_W} (y) .
    \end{split}
\end{equation}
Since the terms accompanying the new variable in the argument of the Hermite polynomials are constant, we can simplify the notation by defining these terms as
\begin{equation}\label{coef. a and b}
    a_i = \sqrt{\dfrac{B_i}{B_W}} \ \ \ \ \text{and} \ \ \ b_i = \sqrt{B_i}(x_{c_W}-x_{c_i}).
\end{equation}
This allows the integral to assume a visually simplified form
\begin{equation} \label{eq.integral}
    I_1 = \dfrac{1}{\sqrt{B_W}} \int dy e^{-y^2} H_{n_1}(a_1y+b_1) H_{n_2}(a_2y+b_2) H_{n_W}(y).
\end{equation}

To proceed toward the final solution of this integral, we must recall that the product of two polynomials yields a polynomial with a degree equal to the sum of their individual highest degrees. Thus, the product of two Hermite polynomials of degrees $n_1$ and $n_2$ will result in a polynomial of degree $n_1 + n_2$,
\begin{equation} \label{eq.poly}
    \mathcal P^{(n_1+n_2)}(y) = H_{n_1}(a_1y+b_1) H_{n_2}(a_2y+b_2).
\end{equation}
By expanding this polynomial in the complete basis of Hermite polynomials with argument $y$, the representation for this product is given by
\begin{equation}\label{poly. Hermite}
    \mathcal P^{(n_1+n_2)} = \sum_{m=0}^{n_1+n_2} C_m H_m(y)
\end{equation}
where the coefficient $C_m$ must be a composition of the constants $a_1$, $a_2$, $b_1$, and $b_2$.

By substituting this expansion into the integral, we obtain a structure closely resembling that of Eq.~\eqref{orth. hermite},
with the sole distinction that one of the Hermite polynomials is no longer restricted to a specific oscillation index value. Hence, we have
\begin{equation}\label{landau level rule}
    \begin{split}
        I_1 =& \ \dfrac{1}{\sqrt{B_W}} \int dy \ e^{-y^2} \sum_{m=0}^{n_1+n_2} C_m H_m(y) H_{n_W}(y) \\
        =& \ \dfrac{1}{\sqrt{B_W}} \sum_{m=0}^{n_1+n_2} C_m  \int dy \ e^{-y^2} H_m(y) H_{n_W}(y)\\
        =& \ \sqrt{\dfrac{\pi}{B_W}} \  \sum_{m=0}^{n_1+n_2}  \Big(C_m \ 2^{n_W} n_W! \delta_{m,n_W} \Big).
    \end{split}
\end{equation}
The physical interpretation of this result establishes that, given the initial conditions for $n_1$ and $n_2$, the generated charged $W$-boson can occupy an oscillator index within the range $n_W \in \{0, 1, \dots, n_1+n_2\}$. This is in perfect agreement with the particular case addressed in Sec.~\ref{secIII},
where the restriction to the lowest oscillator index ($n_1 = n_2 = 0$), coming from choosing the LLLs, resulted in a vanishing sum, thereby restricting the $W^+$ boson solely to the $n_W = 0$ state.  Furthermore, the coefficients $C_m = C_m(a_1, a_2, b_1, b_2)$, which inherently depend on the charges and momenta of the initial particles, act as weight factors. They determine the magnitude of each contribution to the generated vertex and, consequently, dictate which of these possible transitions is the most probable to occur.

The final step in this construction consists simply of assembling all the derived objects, combining both the tensor and spatial parts, and writing out all defined terms explicitly. The resulting general expression for the interaction vertex of two fermions generating a $W$-boson, whether positively or negatively charged, is given by

\begin{equation} \label{selection rules}
    \begin{split}
        \mathcal V =&  \ V_{f_1,f_2}\sum_{\lambda_1,\lambda_2,\lambda_W}\tau^{\lambda_1,\lambda_2, \lambda_W} \dfrac{(4\pi B_1)^{1/4}}{\sqrt{n_1!}} \dfrac{(4\pi B_2)^{1/4}}{\sqrt{n_2!}}  \\
        &\times (4\pi B_W)^{1/4} \ \delta_{023}(p_1+p_2-q_W)\  2^{-\frac{n_1}{2}}2^{-\frac{n_2}{2}}\\
        &\times  e^{-\frac{\Theta}{2}} \sqrt{\dfrac{\pi}{B_W}} \sum_{m=0}^{n_1+n_2}  \Big(C_m \ 2^{\frac{n_W}{2}} \sqrt{n_W!} \ \delta_{m,n_W} \Big)
    \end{split}
\end{equation}
It should be noted that, although $\mathcal V$ depends on the orbit centers $x_{c_i}$ and $x_{c_W}$ (and therefore on the quantum number $q^2$ of the corresponding particles) appearing in $\Theta$ and $C_m$, such dependence
should disappear when this matrix element is used to evaluate a physical process. An example of this can be found in Refs~\cite{Coppola:2018ygv,Coppola:2019wvh} where the matrix elements describing the weak decay of charged pions are evaluated and used to determine the corresponding decay width.

\section{Specific results for some of the next to the lowest Landau level transitions}\label{secV}

Having established the procedure to generalize the selection rules for any process involving two fermions generating a $W^\pm$-boson, we can now test the obtained results through specific examples. Our primary interest lies in demonstrating the possible selection rules for the interaction vertex involving the next to lowest Landau levels. To this end, we first describe how the coefficients $C_m$ must be expressed for these states. We employ the particular case addressed in Sec.~\ref{secIII} to gather these terms. Subsequently, once the $C_m$ coefficients are obtained, we analyze the tensor structure on a case-by-case basis for these Landau levels. The objective is to explicitly reveal the allowed configurations dictated by the selection rules established in Sec.~\ref{secIV} for the more general process involving the quark $u$ and the antiquark $\bar{d}$.

\subsection{The \texorpdfstring{$C_m$}{Cm} coefficients for the next to lowest Landau levels of the \texorpdfstring{$u$}{u} \texorpdfstring{$\bar{d}$}{bar u} \texorpdfstring{$\rightarrow$}{rightarrow} \texorpdfstring{$W^+$}{W+} process}

First, we exemplify transitions involving the next Landau levels for the initial particles generating the boson. To maintain the logical framework established in Sec.~\ref{secIII}, we focus our example on the specific $u\bar{d} \rightarrow W^+$ process, once again assuming that the initial fermions have their spins aligned with the background magnetic field.

This implies that the tensor structure remains exactly the same as previously obtained, regardless of the Landau level occupied by the initial particles. In other words, the generated $W^+$-boson has a definite spin projection along the magnetic field. This does not imply that its polarization vector is parallel to the field; rather, the corresponding circular polarization vector lies in the transverse plane and carries a definite spin projection along the field direction. Consequently, for the following examples, we consider $s_u = s_{\bar{d}} = s_W = +1$, along with $\lambda_u = +1$, $\lambda_{\bar{d}} = -1$, and $\lambda_W = -1$. Thus, we only specify the Landau level of each initial particle to explicitly determine the form of the coefficients $C_m$,
since the permissible states for the boson are already dictated by the oscillator index selection rule Eq.~\eqref{poly. Hermite}.

\begin{itemize} \item[1)] Choosing $k_u = 1$ and $k_{\bar{d}} = 0$: \end{itemize}
The choices $k_u = 1$ and $k_{\bar{d}} = 0$ yield the oscillator indices $n_u = 1$ and $n_{\bar{d}} = 0$, which restrict the generated boson to the range $0 \leq n_W \leq 1$, in accordance with the selection rules derived in Eq.~\eqref{landau level rule}. Thus, employing Eq.~\eqref{eq.poly}, we obtain
\begin{equation}
    \begin{split}
                H_1(a_1y+b_1)H_0(a_2y+b_2) =& \ C_0 H_0(y)+C_1 H_1(y)
    \end{split}
\end{equation}
which, using the explicit form of the Hermite polynomials, implies
\begin{equation}
    \begin{split}
                2\, a_1 \ y+2\, b_1 =& \ C_0+ 2\, C_1 \ y.
    \end{split}
\end{equation}

Comparing the order of each polynomial term, we find that for this first excited oscillator index configuration, the coefficients $C_m$ present in the sum of Eq.~\eqref{selection rules} are explicitly given by
    \begin{align}
         C_1 &= a_1 ,\notag \\
        C_0 &= 2b_1 .
    \end{align}

\begin{itemize}\item[2)] Choosing $k_u = 0$ and $k_{\bar{d}} = 1$:\end{itemize}
The procedure and logic of this second scenario are identical to those of the previous case, and the allowed Landau levels for the production of the $W^+$-boson remain unchanged. The only difference is that the first excited state now corresponds to the $\bar{d}$ antiquark, which exclusively affects the functional form of the coefficients. To make the algebraic development explicit, the polynomial equality reads
\begin{equation}
    \begin{split}
                2a_2y+2b_2 =& \ C_0+C_1(2y) .
    \end{split}
    \end{equation}
Consequently, the coefficients corresponding to the $n_u = 0$ and $n_{\bar{d}} = 1$ case are given by
    \begin{align}
         C_1 &= a_2 ,\notag \\
        C_0 &= 2b_2 .
    \end{align}

\begin{itemize}\item[3)] Choosing $k_u = 1$ and $k_{\bar d} = 1$:\end{itemize}
In this final case, the selection rules governing the allowed Landau levels for the generated boson are modified. Since both the $u$ quark and the $\bar{d}$ antiquark now occupy the first
excited oscillator index value ($n_u = n_{\bar{d}} = 1$), the selection rule establishes that the $W^+$ boson can be generated in the states $n_W \in \{0, 1, 2\}$.
Consequently, the sum in Eq.~\eqref{selection rules} now runs from $0$ to $2$, resulting in three independent $C_m$ coefficients. Employing Eq.~\eqref{eq.poly} once again, we obtain the following polynomial equality
\begin{widetext}
            \begin{equation}
                \begin{split}
                    H_1(a_1y+b_1)H_1(a_2y+b_2) =& \ C_0 H_0(y)+C_1 H_1(y) + C_2 H_2(y)
                \end{split}
            \end{equation}
\end{widetext}
Applying the same procedure as before we determine that the coefficients for this excited state configuration are given by
    \begin{align}
        C_2 &= a_1a_2 ,\notag\\
        C_1 &= 2(a_2b_1+a_1b_2),\notag \\
        C_0 &= 4b_1b_2+2a_1a_2 .
    \end{align}

\subsection{Tensor structure for the next to lowest Landau levels}\label{secVB}

Now that we have derived the explicit forms of the $C_m$ coefficients for the next to lowest Landau levels in the previous subsection, our objective is to present all possible interaction vertices permitted by the selection rules established in this paper. For these examples, we will specify only the initial Landau levels and systematically investigate all allowed configurations for the vertex described by process $u \bar{d} \rightarrow W^+$. Since we do not impose fixed initial spin projections, the tensor structure yields four possibilities.

First, let us analyze the configurations that yield a longitudinal polarization, that is, those leading to the selection rule $\lambda_W = 0$. In this scenario, there are two possible choices: $\lambda_u = \lambda_{\bar{d}} = +1$ and $\lambda_u = \lambda_{\bar{d}} = -1$.

Note that the choice $\lambda_u = \lambda_{\bar{d}} = +1$ yields the following tensor structure
\begin{widetext}
    \begin{equation}\label{tensor ++W}
    \begin{split}
        \tau^{+1,+1, \lambda_W}=& \ \delta_{0,\lambda_W} \left[ \bar{v}_{-Q_2} \ \gamma_{||,\nu} \  \Gamma^{+}  \  \dfrac{(1-\gamma^5)}{2}u_{Q_1}  \right]  \epsilon^{\nu *}_{Q_W}.
    \end{split}
\end{equation}
\end{widetext}
By performing the summation over the $\lambda_W$ indices in Eq.~\eqref{selection rules}, we find that for $\lambda_1 = \lambda_2 = +1$, the Kronecker delta in Eq.~(\ref{tensor ++W}) selects $\lambda_W = 0$. Here $\lambda_W$ labels the Ritus tensor sector: $\lambda_W=0$ corresponds to the longitudinal $(0,3)$ subspace defined relative to the magnetic-field direction, whereas $\lambda_W=\pm 1$ correspond to the transverse $(1,2)$ sector. 
We emphasize that $\lambda_W=0,\pm 1$ labels the tensor components entering the Ritus representation, whereas $c=1,2,3$ labels the polarization vectors. The correspondence between them follows from projecting the polarization vectors onto the longitudinal and transverse tensor sectors.

Conversely, for the choice $\lambda_u = \lambda_{\bar{d}} = -1$, the tensor structure takes the form
\begin{widetext}
    \begin{equation}\label{tensor --W}
    \begin{split}
        \tau^{-1,-1, \lambda_W} =& \  \delta_{0,\lambda_W}  \left[ \bar{v}_{-Q_2} \gamma_{||,\nu} \  \Gamma^{-} \ \dfrac{(1-\gamma^5)}{2}u_{Q_1} \right]  \epsilon^{\nu *}_{Q_W}
    \end{split}
\end{equation}
\end{widetext}
This final case is analogous to the first one, as it also results in the longitudinal polarization for the generated $W$-boson, thereby following all the discussions and constraints previously established in the first scenario.

The remaining two cases correspond to the transverse polarizations, namely, the second term in the parentheses of Eq.\eqref{tensor udW F} which dictates the selection rule $\delta_{\lambda_W} = \pm 1$. In this scenario, there are two possible combinations of $\lambda_u$ and $\lambda_{\bar{d}}$ that prevent the term from vanishing: $(+,-)$ or $(-,+)$.

By choosing the configuration $\lambda_u = +1$ and $\lambda_{\bar{d}} = -1$, the tensor structure reduces to
\begin{widetext}
    \begin{equation}\label{tensor +-W}
    \begin{split}
        \tau^{+1,-1, \lambda_W}=& \  \delta_{-\lambda_W ,+1} \left[ \bar{v}_{-Q_2} \ \gamma_{\perp,\nu} \ \Gamma^{+} \ \dfrac{(1-\gamma^5)}{2}u_{Q_1} \right] \epsilon^{\nu *}_{Q_W}.
    \end{split}
\end{equation}
\end{widetext}
Once again, performing the summation over the $\lambda_W$ indices in Eq.~\eqref{selection rules}, we find that for the case where $\lambda_1 = +1$ and $\lambda_2 = -1$, we obtain a transverse projection for the generated $W$-boson. 

Conversely, for the opposite configuration ($\lambda_u = -1$ and $\lambda_{\bar{d}} = +1$), the tensor structure takes the form
\begin{widetext}
    \begin{equation}\label{tensor -+W}
    \begin{split}
        \tau^{-1,+1, \lambda_W}= & \ \delta_{-\lambda_W ,-1} \left[ \bar{v}_{-Q_2} \ \gamma_{\perp,\nu} \Gamma^{-}  \ \dfrac{(1-\gamma^5)}{2}u_{Q_1} \right] \epsilon^{\nu *}_{Q_W}.
    \end{split}
\end{equation}
\end{widetext}
In this scenario, the surviving term from the summation over the $\lambda_W$ indices will also correspond to a transverse projection for the generated $W^+$ boson. However, in this case, we obtain $\lambda_W = +1$, which, in the special case ($u\bar{d} \rightarrow W^+$), represents the other transverse polarization. 

With this, we have covered all possible combinations arising from the initial spin projection choices for the annihilating quark $u$ and antiquark $\bar d$. However, it is crucial to emphasize a particular detail regarding our notation: the particle and antiparticle follow distinct projection rules, a direct consequence of the definitions established in Eq.~\eqref{def. U and V}. As previously illustrated in the LLL analysis for the $u\bar{d}$ quark-antiquark pair,
to obtain the expected projection in the lowest oscillator index value (both aligned with the background field, since they share the same charge sign, $Q_u,Q_{\bar{d}}> 0$), we had to select $\lambda_u = +1$ and $\lambda_{\bar{d}} = -1$. This implies that for an antiparticle to project its spin in the same direction as a particle, we must adopt the opposite sign for the antiparticle's $\lambda$ index. This equivalence also holds true if we seek a spin state anti-aligned with the magnetic field. While anti-aligning the particle spin requires choosing $\lambda_u = -1$, achieving this exact same physical configuration for the antiparticle will result in selecting $\lambda_{\bar d} = +1$.

\subsection{Vertex results for some of the next to lowest Landau levels}\label{subVC}

Having established the tensor structure and the explicit form of the $C_m$ coefficients for the next to lowest Landau levels, we now construct the allowed interaction vertices for each of the examples previously listed in this
section. It is worth recalling that we are generalizing the analysis of Sec.~\ref{secIII} to scenarios in which we do not fix the initial spin projections of the quark-antiquark pair generating the $W^+$ boson.
Here, we solely specify the Landau levels occupied by these initial particles.
For convenience, we introduce the following shorthand notation
\begin{eqnarray}
\beta = V_{u,d} \ (4\pi B_1)^{1/4} (4\pi B_2)^{1/4} (4\pi B_W)^{1/4}\sqrt{\dfrac{\pi}{B_W}}e^{-\frac{\Theta}{2}}\ \
\end{eqnarray}
where $\Theta$ has been defined in Eq~(\ref{theta}).

\begin{itemize} \item[1)] Choosing $k_u = 1$ and $k_{\bar{d}} = 0$: \end{itemize}
As demonstrated for this scenario, the possible oscillator indices for the generated $W^+$-boson are $n_W = 0$ and $n_W = 1$, which correspond to the Landau levels $k_W = -1$ and $k_W = 0$, respectively.
Consequently, the only allowed vertex combinations will be those whose tensor structures yield physically accessible polarization states for the $W^+$ boson at these specific levels.

If we consider the boson generated in the LLL (and consequently $n_W = 0$), the only permitted polarization state corresponds to $c = 1$. Thus, the single possible interaction vertex is expressed by
\begin{widetext}
\begin{equation}
    \begin{split}
        \mathcal V =& \sqrt2 \beta  b_1  \ \delta_{023}(p_1+p_2-q_W) \ \Bigg[  \bar{v}_{-Q_d} \gamma_{\perp,\nu} \ \Gamma^{+} \dfrac{(1-\gamma^5)}{2}u_{Q_u} \Bigg] \epsilon^{\nu *}_{Q_W}(-1,q^3, c=1)
       .
    \end{split}
\end{equation}
\end{widetext}


On the other hand, for the oscillator index $n_W = 1$, the polarization $c=1$ remains accessible, and $c=2$ also becomes permitted, as can be seen in Eq.~\eqref{polarization k=0}. The latter allows for two distinct configurations, given that both Eq.~\eqref{tensor ++W} and Eq.~\eqref{tensor --W} yield a longitudinal projection $\lambda_W = 0$. 

An interesting point to highlight at this stage is that by observing Eq.~\eqref{polarization k=0}, one can notice that for $k_W = 0$, the polarization vector $c = 1$ does not exhibit vanishing parallel components. That is, when analyzing the $\lambda_W = 0$ case, the polarization vector $c = 1$ should also be a possible choice, alongside the vector $c=2$. Meanwhile, for $\lambda_W = -1$, the only choice will be $c=1$, since the other permitted polarization vector ($c=2$) belongs purely to what we refer to in this text as the longitudinal sector.

Consequently, all possible interaction vertices for this oscillator index are given by
\begin{widetext}
\begin{equation}
    \begin{split}
        [\lambda_u = + , \lambda_{\bar d} = - \ \text{and} \ \lambda_W = - \ ] \rightarrow \ \mathcal V =& \beta \   a_1 \  \delta_{023}(p_1+p_2-q_W) \Bigg[ \bar{v}_{-Q_d} \gamma_{\perp,\nu} \ \Gamma^{+}
        \dfrac{(1-\gamma^5)}{2} u_{Q_u} \Bigg] \ \epsilon^{\nu *}_{Q_W}(0,q^3,c=1)
    \end{split}
\end{equation}
\begin{equation}
    \begin{split}
        &[\lambda_u = \pm , \lambda_{\bar d} = \pm \ \text{and} \ \lambda_W = 0] \rightarrow \ \mathcal V = \beta \  a_1  \delta_{023}(p_1+p_2-q_W) \
        \left[ \bar{v}_{-Q_d} \ \gamma_{||,\nu}    \Gamma^{\pm} \dfrac{(1-\gamma^5)}{2}u_{Q_u} \right] \ \epsilon^{\nu *}_{Q_W}(1,q^3, c),\\
        &\text{with} \ c = 1,2.
    \end{split}
\end{equation}
\end{widetext}


\begin{itemize}\item[2)] Choosing $k_u = 0$ and $k_{\bar{d}} = 1$:\end{itemize}
For this second scenario, the underlying reasoning remains unchanged. Given that the sum of the oscillator indices is $n_u+n_{\bar{d}} = 1$, the accessible Landau levels and the respective permitted polarizations for the $W^+$ boson are strictly the same as in the previous case. The only distinction between the scenarios lies in the functional form of the $C_m$ coefficients, established in the preceding subsection. Therefore, the four allowed interaction vertices are given by:
\begin{widetext}
\noindent  For $n_W = 0$
\noindent \begin{equation}\label{n_w=0 with u=+, d=-, c =1}
    \begin{split}
        [\lambda_u = + , \lambda_{\bar d} = - \ \text{and} \ \lambda_W = - ] \rightarrow \mathcal V =&
         \sqrt{2} \ \beta \ b_2  \ \delta_{023}(p_1+p_2-q_W) \Bigg[ \bar{v}_{-Q_d} \gamma_{\perp,\nu} \, \Gamma^{+} \, \dfrac{(1-\gamma^5)}{2}u_{Q_u} \Bigg] \epsilon^{\nu *}_{Q_W}(-1,q^3,
         c=1).
    \end{split}
\end{equation}
For $n_W = 1$
\begin{equation}\label{n_w=1 with u=+, d=-, c =1}
    \begin{split}
        [\lambda_u = + , \lambda_{\bar d} = - \ \text{and} \ \lambda_W = - ] \rightarrow  \mathcal V =&  \beta \ a_2 \ \delta_{023}(p_1+p_2-q_W) \left[  \bar{v}_{-Q_d}
        \gamma_{\perp,\nu}\Gamma^{+}  \dfrac{(1-\gamma^5)}{2}u_{Q_u}\right] \ \epsilon^{\nu *}_{Q_W}(0,q^3,c=1) ,
    \end{split}
\end{equation}

\begin{equation}\label{n_w=1 with u=+, d=+, c =2}
    \begin{split}
        &[\lambda_u = \pm , \lambda_{\bar d} = \pm \ \text{and} \ \lambda_W = 0] \rightarrow \mathcal V =  \beta \ a_2 \ \delta_{023}(p_1+p_2-q_W) \left[ \bar{v}_{-Q_d} \gamma_{||,\nu}\
        \Gamma^{\pm}\ \dfrac{(1-\gamma^5)}{2}u_{Q_u} \right] \epsilon^{\nu *}_{Q_W}(0,q^3, c) , \\ &\text{with} \ c = 1,2.
    \end{split}
\end{equation}
\end{widetext}
\begin{itemize}\item[3)] Choosing $k_u = 1$ and $k_{\bar d} = 1$:\end{itemize}
In this final case, the selection rule dictates that the $W^+$ boson is restricted to the Landau levels $k_W =-1, 0, 1$, which correspond to the oscillator indices $n_W =0, 1, 2$. To avoid a proliferation of equations, since all possible combinations
would yield eight distinct interaction vertices, we note that the structures for the oscillator indices $n_W = 0$ and $n_W = 1$ remain identical to those obtained in the previous scenarios. The only modification lies in the form of the $C_m$ coefficients present in each term. The required substitution is

\begin{align}\label{subs. C_m}
    C_0 &= 2b_1 \ \text{or} \ 2b_2 \rightarrow 4b_1b_2+2a_1a_2 \notag\\
    C_1 &= a_1 \ \text{or} \ a_2 \rightarrow 2(a_2b_1+a_1b_2)
\end{align}
Thus, for this scenario, the first four vertices assume identical forms to those in Eqs.~\eqref{n_w=0 with u=+, d=-, c =1},~\eqref{n_w=1 with u=+, d=-, c =1} and \eqref{n_w=1 with u=+, d=+, c =2}, requiring only the substitution introduced in Eq.~\eqref{subs. C_m} and the inclusion of an overall factor of $2^{-1}$ for the vertex with $n_W = 0$ and a factor of $2^{-1/2}$ when $n_W = 1$.

It remains for us, therefore, to write down the final four interaction vertices associated with the highest allowed oscillator index for this scenario, $n_W = 2$. This state is particularly interesting because it corresponds to the Landau level $k_W = 1$ for the $W^+$ boson. Consequently, all polarization vectors become accessible, thereby exposing all permitted tensor structures for the vertex. For higher Landau levels, the structure governing the underlying physics of the interaction will remain strictly the same, experiencing only modifications to the $C_m$ coefficients. Therefore, we conclude that the remaining possible vertices for this configuration take the form
\begin{widetext}

    \noindent \begin{equation}
    \begin{split}
        &[\lambda_u = \pm , \lambda_{\bar d} = \pm \ \text{and} \ \lambda_W = 0] \rightarrow  \mathcal V =
        \ \beta \ a_1 \, a_2 \ \delta_{023}(p_1+p_2-q_W) \left[ \bar{v}_{-Q_d} \ \gamma_{||,\nu} \  \Gamma^{\pm} \ \dfrac{(1-\gamma^5)}{2}u_{Q_u}\right]  \ \epsilon^{\nu *}_{Q_W}(1,q^3, c), \\
        &\text{with} \ c = 1,2,3.
    \end{split}
\end{equation}

\begin{equation}
    \begin{split}
        &[\lambda_u = \pm , \lambda_{\bar d} = \mp \ \text{and} \ \lambda_W = \mp \ ] \rightarrow \mathcal V =  \beta \ a_1 \, a_2 \ \delta_{023}(p_1+p_2-q_W)
        \left[ \bar{v}_{-Q_d} \gamma_{\perp,\nu}\, \Gamma^{\pm}\, \dfrac{(1-\gamma^5)}{2}u_{Q_u}\right] \ \epsilon^{\nu *}_{Q_W}(1,q^3,c),\\
        &\text{with} \ c = 1,3.
    \end{split}
\end{equation}
\end{widetext}
\subsection{A brief analysis of the \texorpdfstring{$d \bar{u} \rightarrow W^-$}{W} process}\label{secVD}

For the production of a $W^-$ boson from the annihilation of the $d \bar{u}$ quark-antiquark, the procedure follows exactly the same line of reasoning employed throughout this section. To determine the selection rules, we employ Eq.~\eqref{selection rules} and maintain the exact same structure for the $C_m$ coefficients, requiring only the substitution of the appropriate definitions when using Eq.~\eqref{coef. a and b}.

The most crucial aspect of this procedure is verifying the $\lambda_d$ and $\lambda_{\bar{u}}$ indices in the LLL. Since we now have $Q_d, Q_{\bar{u}} < 0$, theory dictates that in the ground Landau state, the spins will be anti-aligned with the background magnetic field. This implies that we will have $\lambda_d = -1$ and $\lambda_{\bar{u}} = +1$, as discussed at the end of Sec.~\ref{secVB}. It is known that in the LLL, the only permitted polarization vector for the boson must be the one mentioned in Eq.~(\ref{polarization k=-1}), and for the $W^-$ case, this polarization must correspond to anti-alignment with the magnetic field. Observing the $\lambda_W$ value generated in this configuration by the selection rules established in this work, we obtain $\lambda_W = +1$. 
Therefore, unlike the $W^+$ case, where $\lambda_W = -1$ was allowed from the lowest level $k_W \geq -1$, this projection will now have an associated polarization vector only from $k_W \geq 1$. Thus, if one is interested in deriving the possible vertices for the $d\bar{u} \rightarrow W^-$ process, a careful analysis will be required to establish the correct interpretation of the polarizations in this particular scenario.

\section{Summary and conclusions}\label{concl}
In this work, we have derived the selection rules and the explicit structure of the lowest-order interaction vertex between a quark, an antiquark, and charged $W^{\pm}$-bosons in the presence of an external magnetic field. By employing the Ritus eigenfunction method, we expanded the fermion and vector boson fields into Landau levels, which intrinsically account for the anisotropic dynamics induced by the background field. Our analysis of the space-time and tensor structures revealed strict physical constraints on the allowed states of the generated $W$-boson.

For the simplest configuration, where the initial quark and antiquark occupy the LLL with their spins aligned with the magnetic field,
the selection rules dictate that the produced $W$-boson must also occupy its respective LLL (and consequently $n_W=0$) and strictly adopt a transverse polarization aligned with the external field.
We subsequently generalized this formalism to arbitrary initial oscillator indices, $n_1$ and $n_2$. From the spatial overlap of the corresponding Hermite polynomials,
we demonstrated that the allowed oscillator indices for the $W$-boson are restricted to a finite range given by $0 \le n_W \le n_1 + n_2$.

To validate our framework, we explicitly computed the $C_m$ weight coefficients for the next-to-lowest Landau level transitions and detailed the resulting polarization states. Finally, a brief analysis of the $d\bar{u} \rightarrow W^-$ process highlighted that, due to the negative charges of the initial particles, the spin anti-alignment with the magnetic field in the LLL reverses the respective longitudinal and transverse polarization projections compared to the $W^+$ case. These findings provide a solid theoretical foundation for future rigorous calculations of electroweak processes in strongly magnetized environments.
\\

\section*{Acknowledgments}

Support for this work has been received in part by DGAPA-PAPIIT-UNAM grant numbers AG100826 and IN116326, and from Secretar\'\i a de Ciencia, Humanidades, Tecnolog\'\i a e Innovaci\'on (SECIHTI) M\'exico grant numbers CIORGANISMOS-2025-17, CBF-2025-G-1718, and CBF-2026-465, and fellowship number 1184435. This study was financed in part by the São Paulo Research Foundation (FAPESP) Brazil, Process Numbers 2023/08826-7, 2024/18493-8 and 2026/03513-9.

\vspace*{1cm}

{\appendix

\section{Polarization selection rules}\label{appA}

To fully describe the $W$-boson, one must choose a complete polarization vector set $\epsilon_{\mathcal{Q},\nu}(k, q^3, c)$, with $c$ being the index denoting the polarization state, as mentioned in Sec.~\ref{secII}. For $k \geq 1$ there are three linearly independent vectors that satisfy the transversality condition \cite{GomezDumm:2023owj}

\begin{equation} \label{eq.transver}
\Pi^{\mu}(k, q_{\parallel})^{*}\Big|_{q^{0} = E_{W}} \epsilon_{\mathcal{Q}, \mu}(k, q^{3}, c) = 0,
\end{equation}

where

\begin{equation}
\begin{split}
\Pi^{\mu}(k, q_{\parallel}) & \equiv \Big( q^{0}, i \sqrt{B_{W}/2} \left( \sqrt{k + 1} - \sqrt{k} \right), \\
& \hspace{0.5cm} - s_{W} \sqrt{B_{W}/2} \left( \sqrt{k + 1} + \sqrt{k} \right), q^{3} \Big),
\end{split}
\end{equation}
is a four-vector that, in some cases, plays a role equivalent to the on played by the four-momentum vector for $B = 0$. A convenient choice of polarization vectors that fulfill Eq. (\ref{eq.transver}) for $k \geq 1$ is

\begin{widetext}

\begin{equation}\label{polarization k>=1}
\begin{split}
\epsilon^{\mu}_{\mathcal{Q}}(k, q^{3}, 1) & = \frac{1}{\sqrt{2}} \frac{1}{m_{\bot} m_{2 \bot}} \left[ \Pi_{+} \left( E_{W}, 0, 0, q^{3} \right) + m_{\bot}^{2} \left( 0, 1, is_{W}, 0 \right) \right], \\
\epsilon^{\mu}_{\mathcal{Q}}(k, q^{3}, 2) & = \frac{1}{m_{\bot}} \left( q^{3}, 0, 0, E_{W} \right), \\
\epsilon^{\mu}_{\mathcal{Q}}(k, q^{3}, 3) & = \frac{1}{\sqrt{2}} \frac{1}{m_{W} m_{2 \bot}} \left[ \Pi_{-} \left( E_{W}, \frac{\Pi^{*}_{+}}{2}, i s_{W} \frac{\Pi^{*}_{+}}{2}, q^{3} \right) + m^{2}_{2 \bot} \left( 0, 1, -is_{W}, 0 \right) \right],
\end{split}
\end{equation}
\end{widetext}
where we have used the definitions
\begin{equation}
\begin{split}
m_{\bot} & = \sqrt{m_{W}^{2} + \left( 2k + 1 \right) B_{W}}, \\
m_{2 \bot} & = \sqrt{m_{W}^{2} + k B_{W}}, \\
\Pi_{\pm} & = -\Pi^{1}(k, q_{\parallel}) \pm is_{W} \Pi^{2}(k, q_{\parallel}).
\end{split}
\end{equation}

For $k = 0$ only two independent nontrivial polarization vectors can be constructed. An appropriate choice of polarization vectors is

\begin{equation}\label{polarization k=0}
\begin{split}
\epsilon^{\mu}_{\mathcal{Q}}\left( 0, q^{3}, 1 \right) & = \frac{1}{\sqrt{2}} \frac{1}{m_{\bot} m_{2 \bot}} \left( E_{W} \Pi_{+}, m_{\bot}^{2}, i s_{W} m^{2}_{\bot}, q^{3} \Pi_{+} \right) \\
\epsilon^{\mu}_{\mathcal{Q}}\left( 0, q^{3}, 2 \right) & = \frac{1}{m_{\bot}} \left( q^{3}, 0, 0, E_{W} \right),
\end{split}
\end{equation}
where $m_{\bot}, m_{2 \bot}, \Pi_{+}$ and $E_{W}$ are understood to be evaluated at $k = 0$.

In the case of the LLL, for $k = -1$, one has that $\mathbb{R}^{\mathcal{Q},\mu\nu}(x, \bar{q}) \propto \Upsilon^{\mu \nu}_{-s_{W}}$. In this case, only one nontrivial polarization vector can be defined, a convenient choice of such polarization vector is

\begin{equation}\label{polarization k=-1}
\epsilon^{\mu}_{\mathcal{Q}}\left( -1, q^{3}, 1 \right) = \frac{1}{\sqrt{2}} \left( 0, 1, is_{W}, 0 \right).
\end{equation}
As in the case of $\epsilon^{\mu}_{\mathcal{Q}}\left( 0, q^{3}, 2 \right)$, it is easy to see that this vector has a definite spin projection in the direction of the magnetic field. Indeed, one has

\begin{equation}
S^{\mu \nu}_{3} \epsilon_{\mathcal{Q}, \nu}(-1,q^{3},1) = s_{W} \epsilon^{\mu}_{\mathcal{Q}}(-1,q^{3},1),
\label{wspin}
\end{equation}
see Eq.~(C.17) of Ref.~\cite{GomezDumm:2023owj}.

\section{General form of the coefficients of the expansion of Hermite polynomials}\label{appB}

To obtain the coefficients mentioned in Sec.~\ref{secIVB}, a general solution for the integral in Eq.~(\ref{eq.integral}) is required. Such integral can be rewritten as
\begin{equation}
\begin{split}
I_{1} & = \frac{1}{\sqrt{B_{W}}} \sum_{m_{1} = 0}^{n_{1}} \sum_{m_{2} = 0}^{n_{2}} \mathcal{K}_{1}\left( n_{1}, m_{1} \right)  \mathcal{K}_{2} \left( n_{2}, m_{2} \right) \\
&  \times   \int dy \; e^{-y^{2}}  H_{m_{1}} \left( a_{1} y \right) H_{m_{2}} \left( a_{2} y \right) H_{n_{W}}(y)
\end{split}
\end{equation}
where the known result~\cite{morse1953methods}
\begin{equation}
H_{n} (z + b) = \sum_{m = 0}^{n} \frac{\left( 2 b \right)^{n - m} n!}{m! \left( n - m \right)!} H_{m} (z),
\end{equation}
has been used and the definition
\begin{equation}
\mathcal{K}_{i}(n_{i}, m_{i}) \equiv \frac{\left( 2 b_{i} \right)^{n_{i} - m_{i}} n_{i}!}{m_{i}!\left( n_{i} - m_{i} \right)!},
\end{equation}
with $i = \{ 1, 2 \}$ has been implemented. Then, after using the rescaling identity
\begin{equation}
H_{n}\left( \lambda z\right) = \sum_{m = 0}^{\lfloor \frac{n}{2} \rfloor} n! \frac{\lambda^{n - 2m} \left( \lambda^{2} - 1 \right)^{m}}{m! \left( n - 2m \right)!} H_{n - 2m}(z),
\end{equation}
the integral can be rewritten as
\begin{equation}
\begin{split}
I_{1} & = \frac{1}{\sqrt{B_{W}}} \sum_{m_{1} = 0}^{n_{1}} \sum_{m_{2} = 0}^{n_{2}} \mathcal{K}_{1}\left( n_{1}, m_{1} \right)  \mathcal{K}_{2} \left( n_{2}, m_{2} \right)\\
    & \times \sum_{r_{1} = 0}^{\lfloor \frac{m_{1}}{2} \rfloor} \sum_{r_{2} = 0}^{\lfloor \frac{m_{2}}{2} \rfloor} \mathcal{R}_{1} \left( m_{1}, r_{1} \right) \mathcal{R}_{2} \left( m_{2}, r_{2} \right) \\
& \times \int dy \; e^{-y^{2}} H_{m_{1 - 2r_{1}}} \left(  y \right) H_{m_{2 - 2r_{2}}} \left(  y \right) H_{n_{W}}(y),
\end{split}
\end{equation}

where the definition

\begin{equation}
\mathcal{R}_{i} \left( m_{i}, r_{i} \right) \equiv m_{i}! \frac{a_{i}^{m_{i} - 2 r_{i}} \left( a_{i}^{2} - 1 \right)^{r_{i}}}{r_{i}! \left( m_{i} - 2 r_{i} \right)!}
\end{equation}

has been used. Finally, after using the known result~\cite{andrews1999special}

\begin{eqnarray}
& &\int dx \; e^{-x^{2}} H_{\ell}(x) H_{m}(x) H_{n}(x)\nonumber\\
&=& \frac{ 2^{\left( \ell + m + n \right) / 2} \;  \ell! m! n! \sqrt{\pi}}{\left( \frac{\ell + m - n}{2} \right)! \left( \frac{m + n - \ell}{2} \right)! \left( \frac{n + \ell - m}{2} \right)!}
\end{eqnarray}
when $\ell + m + n$ is even and the sum of any two of $\ell, m, n$ is not smaller than the third, this last condition can be implemented as a product of Heaviside step functions so that

\begin{equation}
\begin{split}
I_{1} & = \frac{1}{\sqrt{B_{W}}} \mathbb{C}\left( n_{1}, n_{2}, n_{W} \right),
\end{split}
\end{equation}

where

\begin{widetext}

\begin{equation} \label{eq.SolGenCoef}
\begin{split}
\mathbb{C} \left( n_{1}, n_{2}, n_{W} \right) & = \prod_{i = 1}^{2} \sum_{m_{i} = 0}^{n_{i}} \mathcal{K}_{i} \left( n_{i}, m_{i} \right) \sum_{r_{i} = 0}^{\lfloor \frac{m_{i}}{2} \rfloor} \mathcal{R}_{i} \left( m_{i}, r_{i} \right) 2^{\left( m_{1} - 2 r_{1} + m_{2} - 2 r_{2} + n_{W} \right)/2} \\
& \times \frac{ \left( m_{1} - 2 r_{1} \right)! \left( m_{2} - 2 r_{2} \right)! n_{W}! \sqrt{\pi}}{\left( \frac{ m_{1} - 2r_{1} + m_{2} - 2r_{2} - n_{W}}{2} \right)! \left( \frac{ m_{2} - 2r_{2} + n_{W} - \left( m_{1} - 2r_{1} \right)}{2} \right)! \left( \frac{ n_{W} + m_{1} - 2r_{1} - \left( m_{2} - 2r_{2} \right)}{2} \right)!} \\
& \times \theta\left( m_{1} - 2 r_{1} + m_{2} - 2 r_{2} - n_{W} \right) \theta\left( m_{2} - 2 r_{2} + n_{W} - \left( m_{1} - 2 r_{1} \right) \right) \theta\left( n_{W} +  m_{1} - 2 r_{1} - \left( m_{2} - 2 r_{2} \right) \right).
\end{split}
\end{equation}

\end{widetext}

Using this methodology, one can note that the relation
\begin{equation}
\mathbb{C} \left( n_{1}, n_{2}, n_{W} \right) = \sum_{m = 0}^{n_{1} + n_{2}} C_{m} \  2^{\frac{n_{W}}{2}} \sqrt{\pi} \;  n_{W}!  \delta_{m, n_{W}}
\end{equation}
holds after evaluating the coefficient. We recall that $C_m$ has been introduced in Eq.~(\ref{poly. Hermite}).

It is straightforward to see that evaluating the lowest Landau level $(k_{1} = 0 = k_{2})$ and the next to lowest Landau levels ($k_{1} = 0, k_{2} = 1$ \& $k_{1} = 0, k_{2} = 1$) in Eq.~(\ref{eq.SolGenCoef}) gives the same results as the ones discussed in Sec.~\ref{secV}. This is shown in the next subsection.

\subsection{Expansion coefficients of Hermite polynomials for the LLL and next to LLL}

\underline{For the LLL} the choices $k_{1} = 0 = k_{2}$ are made and imply that $n_{1} = 0$ for $\lambda_{1} = 1$ and $n_{2} = 0$ for $\lambda_{2} = -1$. Then, the coefficient is evaluated in $n_{1} = 0 = n_{2}$. The selection rules dictated by the Heaviside step-functions indicate that also $n_{W} = 0$ so that the corresponding coefficient for the LLL configuration is
\begin{equation}
\mathbb{C}(0,0,0) = \sqrt{\pi}.
\end{equation}

\underline{For the next to LLL}, in particular the configuration choice of $k_{1} = 0$ and $k_{2} = 1$, implies that $n_{1} = 0$ for $\lambda_{1} = 1$, and $n_{2} = \{0, 1\}$ if $\lambda_{2} = \{ 1, -1 \}$ respectively. Then, the corresponding coefficients for such combinations are evaluated in $n_{1} = 0, n_{2} = 0$ and $n_{1} = 0, n_{2} = 1$. The selection rules indicated by the Heaviside step-functions for each case can be found by evaluating the following

\begin{widetext}

\begin{equation}
\begin{split}
\mathbb{C}\left( n_{1} = 0, n_{2} = 0, n_{W} = k_{W} \right) & \propto \theta\left( -k_{W} \right) \theta\left( k_{W} \right) \theta\left( k_{W} \right) \implies k_{W} = 0. \\
\mathbb{C}\left( n_{1} = 1, n_{2} = 0, n_{W} = k_{W} + 1 \right) & \propto  \theta \left( - 1 - k_{W} \right) \theta\left( 1 + k_{W} \right) \theta \left( 1 + k_{W} \right) + \theta \left( - k_{W} \right) \theta\left( k_{W} \right) \theta \left( 2 + k_{W} \right) \\\implies k_{W} = \{ -1, 0 \}.
\end{split}
\end{equation}

\end{widetext}
Then, for the longitudinal component, one obtains a coefficient
\begin{equation}
\mathbb{C}\left( 0, 0, 0 \right) = \sqrt{\pi}.
\end{equation}

For the transverse component, one obtains the coefficients
\begin{eqnarray}
\mathbb{C}\left( 0, 1, 0 \right) & =&  2 \sqrt{\pi} b_{2}\nonumber \\
\mathbb{C}\left( 0, 1, 1 \right) & = &\sqrt{2\pi} a_{2},
\end{eqnarray}
Likewise, for the configuration $k_{1} = 1, k_{2} = 0$ the relevant coefficients are
\begin{equation}
\begin{split}
\mathbb{C}\left( 0, 0, 0 \right) & = \sqrt{\pi}, \\
\mathbb{C}\left( 1, 0, 0 \right) & = 2 \sqrt{\pi} b_{1}, \\
\mathbb{C}\left( 1, 0, 1 \right) & = \sqrt{2 \pi} a_{1}.
\end{split}
\end{equation}
}
\bibliography{bibliographyNNS2}{}
\bibliographystyle{apsrev4-1}

\end{document}